\documentclass[epj]{svjour}
\usepackage{graphicx}
\usepackage{epstopdf}
\usepackage{amssymb}
\usepackage{color}
\begin{document}
\title{ANDERSON LOCALISATION IN SPIN CHAINS FOR PERFECT STATE TRANSFER}

\author{R. Ronke\inst{1}, M.P. Estarellas\inst{2}, I. D'Amico\inst{2}, T.P. Spiller\inst{2} and T. Miyadera\inst{3}
}                     

\institute{RR contributed to the paper while at York, but she is not affiliated with the University of York at the point of publishing of the paper \and Department of Physics, University of York, York, YO10 5DD, United Kingdom \and Department of Nuclear Engineering, Kyoto University, Nishikyo-ku, Kyoto, 615-8540, Japan}
\date{Published in EPJD DOI: 10.1140/epjd/e2016-60665-0}

\abstract{
Anderson localisation is an important phenomenon arising in many areas of physics, and here we explore it in the context of quantum information devices. Finite dimensional spin chains have been demonstrated to be important devices for quantum information transport, and in particular can be engineered to allow for `perfect state transfer' (PST). Here we present extensive investigations of  disordered PST spin chains, demonstrating spatial localisation and transport retardation effects, and relate these effects to conventional Anderson localisation. We provide thresholds for Anderson localisation in these finite quantum information systems for both the spatial and the transport domains. Finally, we consider the effect of disorder on the eigenstates and energy spectrum of our Hamiltonian, where results support our conclusions on the presence of Anderson localisation.
\PACS{
      {72.20.Ee}{Mobility edges: hopping transport}   \and
      {03.67.-a}{Quantum Information} \and
      {71.23.An}{Theories and models: localised states} \and
      {75.10.Pq}{Spin chain models}
     } 
} 
\maketitle
\section{Introduction}
\label{intro}
Anderson localisation was predicted in 1958 \cite{anderson1958}, to explain experimental findings of anomalously long relaxation of spins in semiconductors, and then linked to the metal-insulator transition. Since then, its reach and influence has been greatly extended, to many systems and phenomena. Examples include the integer quantum Hall effect \cite{pruisken1988}, classical waves \cite{economou1989,sheng1986}, light diffusion in gallium arsenide or titania powders \cite{wiersma1997,stoerzer2006}, conductance of microwaves in thick wires \cite{chabanov2000}, ultrasound \cite{hu2008}, photonics \cite{segev2013,crespi2013}, cold atomic gases \cite{chabe2008}, and Bose-Einstein condensates \cite{Kondov2011,Jendrzejewski2012}.

Here we focus on one dimensional finite spin chains, which have been set up for `perfect state transfer' (PST) \cite{christandl2004,chris2005,plenio2004}. In recent years this type of spin chain has acquired growing importance within the field of quantum information processing, as a means of efficiently transferring information \cite{bose2007,kay2010}, or for creating and distributing entanglement \cite{spiller2007} within a solid state-based quantum processor or computer. Such chains represent the `perfect wire' for quantum data transmission and, as such, it represents the ideal scenario to study transport deterioration by Anderson localisation. By tuning the couplings, their ideal transport property can be engineered to operate independently from their chain length. However, based on detailed studies of modest length chains it can be conjectured that long chains would be the most affected by random fabrication defects \cite{cai2006,ronke2011_1,ronke2011_2,zwick2011,zwick2015,zwick2015b,ashhab2015} or by slowly varying external fields.
Within this context, we have investigated the question of if and how random defects would affect the relevant transmission properties of PST spin chains, through the appearance of weak localisation or Anderson localisation.
In one dimension, the theory of Anderson localisation \cite{anderson1958} predicts localisation of quantum states under certain conditions. However it should be noted that in one dimensional finite systems disorder does not necessarily imply the complete vanishing of extended states, as discussed by Pendry in the case of `necklace' states \cite{pendry1994}. Therefore there are important theoretical  motivations for considering finite systems, along with the fact that experimental systems are finite in extent.

In this work we consider disordered finite PST spin chains.  Some studies for {\it non}-PST spin chains (i.e. with uniform couplings) have been performed, including the work by  Siber \cite{siber2006} on a single spin chain realization with very strong disorder.
Here we will consider both the transport properties and the spatial localisation of quantum states in finite disordered spin chains, engineered for the important quantum information transport condition of PST. We investigate both a large range of chain lengths and ensembles of random disorder. Our results demonstrate that, due to their properties, PST spin chains may display, for a given level of disorder and within the same chain, different regimes of transport and localisation behaviour (including Anderson localisation). This demonstrates that PST spin chains form an interesting new class of systems in which to study localisation phenomena.

\section{Properties of unperturbed PST spin chains}
\label{sec:1}
The natural dynamics of an $N$-site spin chain, including disorder, can be described by a time independent Hamiltonian as follows

\begin{eqnarray}
\label{hami}
\nonumber{\cal{H}} = \sum_{i=1}^{N}\epsilon_{i}|1\rangle \langle 1|_{i} + \sum_{i=1}^{N-1} J_{i,i+1}[ |1\rangle \langle 0|_{i} \otimes |0\rangle \langle 1|_{i+1} + \\ 
|0\rangle \langle 1|_{i} \otimes |1\rangle \langle 0|_{i+1}]
\end{eqnarray}

  Within spin chains, a single excitation $|1\rangle_i$ at site $i$  is defined as an `up' spin in a system that is otherwise prepared to have all spins in the `down' $|0\rangle$ state.
 For the PST system presented here the single excitation energies $\epsilon_{i}$ are independent of the site $i$, with deviations from this condition being due to disorder, or errors. The coupling strengths $J_{i,i+1}$  between two neighbouring sites $i$ and $i+1$ are pre-engineered as \cite{niko2004,chris2005}
 
\begin{equation}
	J_{i,i+1}=J_{0}\sqrt{i(N-i)}.\label{PST}
\end{equation}

In any practical system there will be a maximum spin-spin coupling strength, independent of the length $N$ and set by a characteristic value for the particular physical realisation of the spins. Therefore, to address this practical constraint, here we keep the maximum coupling value $J_{max}=1$ constant as $N$ is varied. $J_{max}$ is then our unit of energy. The coupling $J_{max}$ occurs in the middle of the chain. As a result $J_{0}=2J_{max}/N$ for even ($J_{0}=2J_{max}/N\sqrt{1-1/N^{2}}$ for odd) length chains.

A useful assessment of chain performance is the fidelity $F$, corresponding to mapping an initial state $|\psi_{ini}\rangle$ over a time $t$ into a desired state $|\psi_{fin}\rangle$, by means of the chain natural dynamics. This is given by

\begin{equation}
	F=|\langle \psi_{fin} |e^{-i{\cal{H}}t/\hbar}| \psi_{ini} \rangle|^{2}
\end{equation}

and PST is realised when the evolution is arranged to achieve $F=1$. We use the fidelity of state vectors to determine the information transport quality. For the PST systems considered here we are interested in the behaviour of a single spin excitation, so we restrict to the single excitation subspace of the chain. For this case, the time scale for an excitation to exhibit PST from one end of a chain to the other is $t_{M}=\pi \hbar/2J_{0}$. More generally, this is the time for any chain state to evolve to its mirror image (about the chain mid-point), so it is also known as \textit{mirroring time} and it scales with $N$ when practical $J_{max}$ is used. In all cases the full periodicity of the system evolution is given by $t_{S}= 2 t_{M}$.
This mirroring phenomenon arises from the fact that for the particular coupling condition Eq.~(\ref{PST}) the spin chain can be mapped onto a macroscopic spin, with the mirroring corresponding to its precession \cite{niko2004,chris2005}. Hence, we can operationally define the mirror operator $M$ having the following effects to each term in any arbitrary superposition state of the chain:
	
\begin{equation}
M|a \rangle_{1} |b\rangle_{2}...|y\rangle_{N-1} |z\rangle_{N}=|z\rangle_{1}|y\rangle_{2}...|b\rangle_{N-1}|a\rangle_{N}
\end{equation}

Any initial state $|\Psi(0)\rangle$ can thus be decomposed into its even and odd parts under the mirror operator $M$, such that

\begin{equation}
|\Psi(0)\rangle=\frac{1}{\sqrt{2}}(|\Psi_{+}(0)\rangle+|\Psi_{-}(0)\rangle)
\end{equation}

being $|\Psi_{\pm}(0)\rangle\equiv\frac{1}{\sqrt{2}}(|\Psi(0)\rangle \pm M|\Psi(0)\rangle)$. Thus the Hamiltonian eigenstates (which are also eigenstates of $M$) can be decomposed as superpositions of even and odd energy eigenstates $|\Psi_{\pm}(0)\rangle\equiv\sum_{\pm\kappa}c_{\pm\kappa}|E_{\pm\kappa}\rangle$. Then for the evolved state at time $t_{M}$ to have unit fidelity against the mirrored initial state $M|\Psi(0)\rangle$, it has to be of the form

\begin{equation}
|\Psi(t_{M})\rangle=\frac{e^{-i\theta}}{\sqrt{2}}(\sum_{+\kappa}c_{+\kappa}|E_{+\kappa}\rangle-\sum_{-\kappa}c_{-\kappa}|E_{-\kappa}\rangle)
\label{evol}
\end{equation}

It is therefore clear that this mirroring phenomenon arises due to the properties of the eigenstates and the eigenvalue spectrum (see l.h.s of Fig.\ref{levels} in our later discussions) that appear from the particular coupling condition, in such a way that the phases in the evolved state conspire to give the form (\ref{evol}) at the mirroring time $t_{M}$.

\begin{figure*}
\resizebox{1\textwidth}{!}{%
  \includegraphics{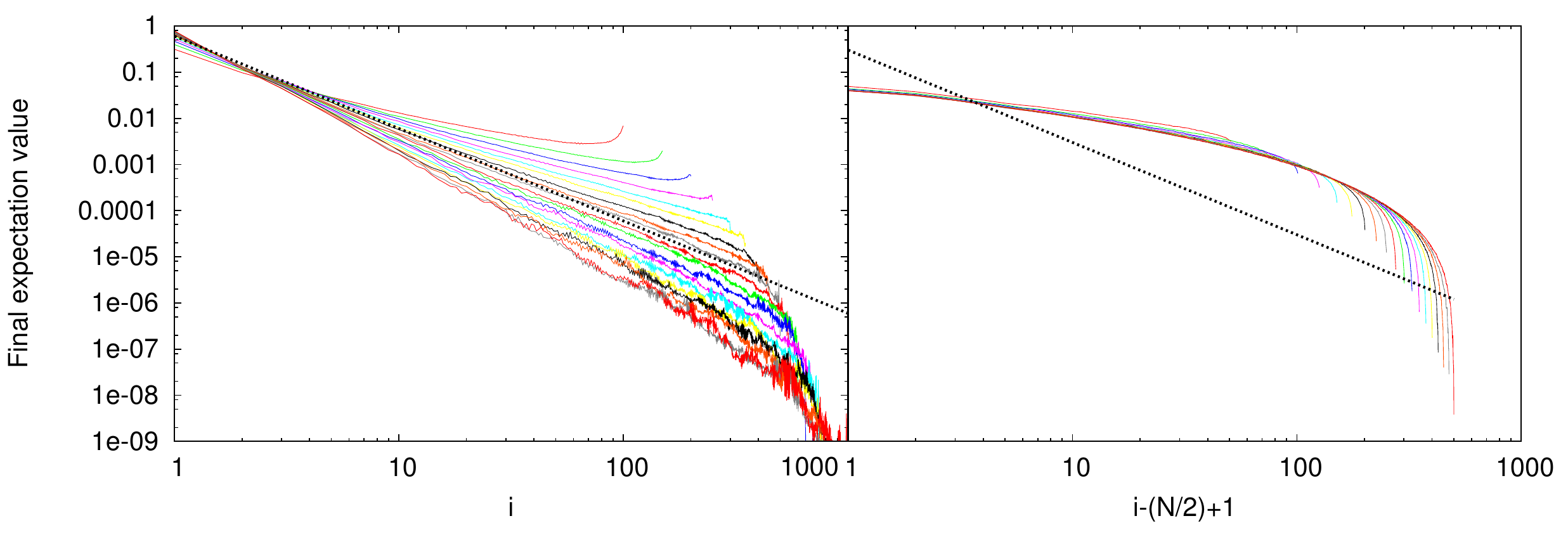}
}
\caption{Effective steady state site occupation probabilities versus chain site, for a range of chain lengths $N$ from $100$ to $1000$ and disorder of strength $E=1.0$, when the excitation is injected at site $1$ (left panel) and $\frac{N}{2}$ (right panel). For comparison, critical lines given by $i^{-2}$ (left) $[i-(N/2)+1]^{-2}$ (right) are added to both plots (with a normalisation factor $0.6$ on the left and a factor $0.3$ on the right due to the double-sided nature of this distribution). These lines give the accepted cut-off for Anderson localisation in the limit $N \rightarrow \infty$. Identifying everything below the cut-off as localised, and everything above as not, we conclude that chains with $N$ about $\sim 500$ and above exhibit Anderson localisation for an excitation initially at the chain end, while chains with the excitation initially centred do not attain this condition for any $N$ in the range investigated.}
\label{endtime2D}       
\end{figure*}

\begin{figure*}
\resizebox{1\textwidth}{!}{%
  \includegraphics{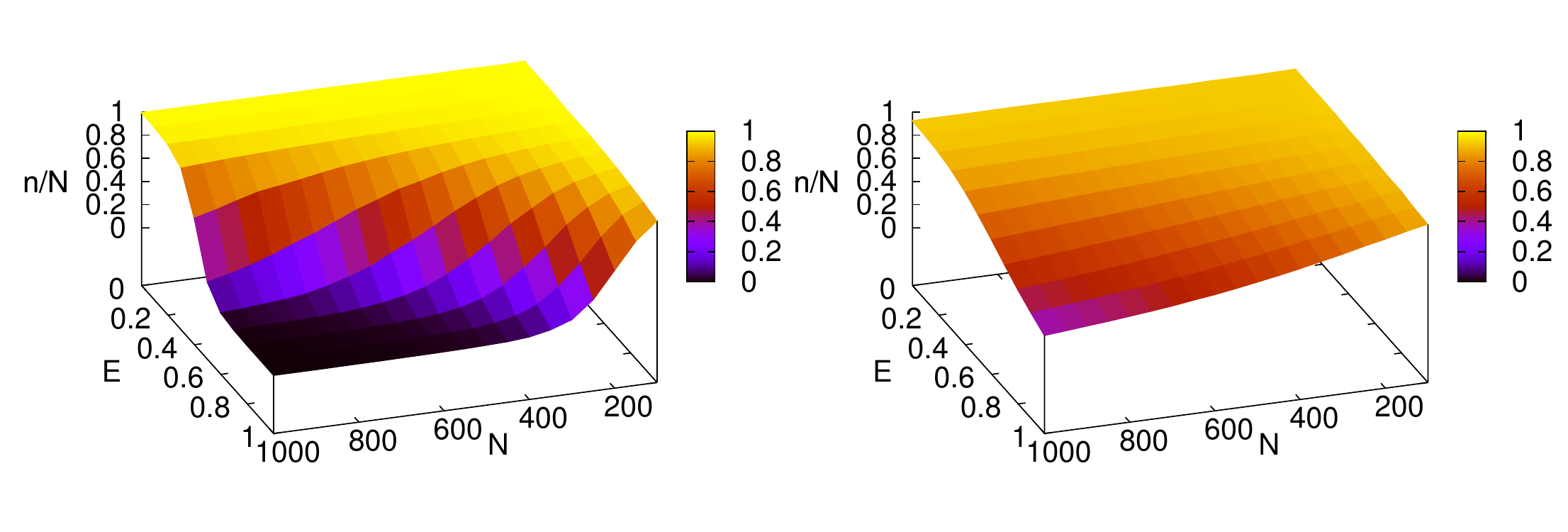}
}
\caption{Ratio $n/N$ vs. chain length $N$ and perturbation strength $E$ for achieving a total site occupation probability of $0.95$ (see Eq.~(\ref{n})), when the excitation is injected at site $i=1$ (left panel) and $i=\frac{N}{2}$ (right panel).}
\label{width}     
\end{figure*}

\begin{figure}[b]
	\resizebox{0.5\textwidth}{!}{%
		\includegraphics{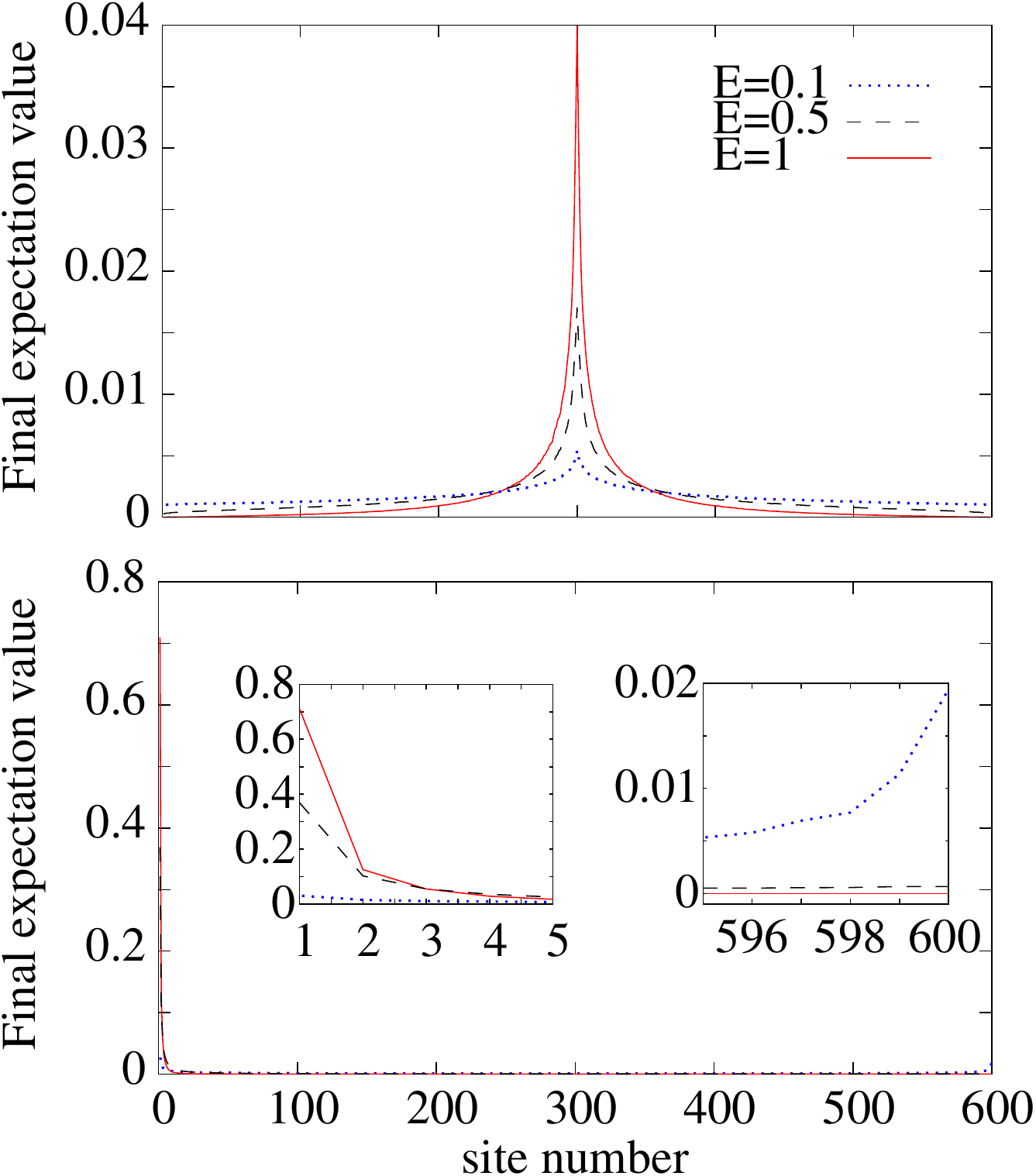}
	}
	\caption{Steady state site occupation probabilities versus site number $i$, shown for three selected disorder strengths $E=0.1$ (dotted line), $E=0.5$ (dashed line), and $E= 1.0$ (solid line), for a chain of length $N=600$, when the excitation is injected at site $i=1$ (bottom panel) and $i=\frac{N}{2}$ (top panel). The insets in the bottom panel focus on the initial (left inset) and final (right inset) sites of the chain.  }
	\label{endtime}   
\end{figure}

\begin{figure} [t]
	\resizebox{0.5\textwidth}{!}{%
		\includegraphics{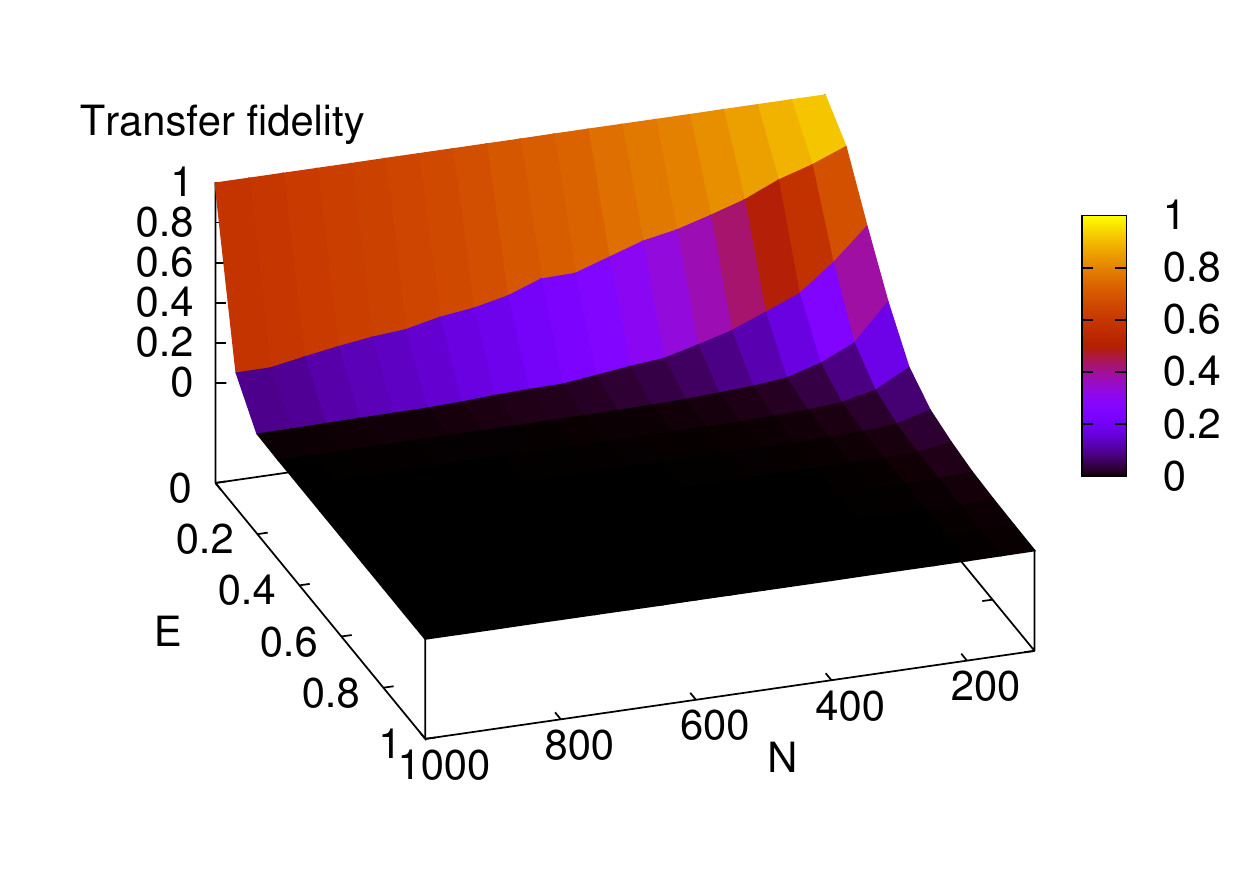}
	}
	\caption{Maximum fidelity of the transferred state in a window of $4.5t_M$ versus chain length $N$ and perturbation strength $E$, for excitation released at the chain end site $i=1$. Results are averaged over 100 disorder realizations.}
	\label{avg_max_3D}      
\end{figure}

\section{Disorder and Anderson localisation}
\label{sec:2}
To simulate practical diagonal disorder in chains with different $N$, we fix the scale of the disorder by $J_{max}$ and set $\epsilon_{i} = E J_{max} d_{i}$, where $d_{i}$ is a random number from a uniform distribution within 0 and 1, and $E$ is a dimensionless parameter that sets the scale of the disorder.
In what follows we will consider the dynamics of a single excitation, injected either at the beginning or in the middle  of the chain. The PST type of chain we analyse in this paper, when unperturbed, ensures perfect state transfer (PST) not only between its end spins, but also between any pair of spins at opposite but equal distance with respect to the chain centre due to the `mirroring' property introduced in the previous section. Within quantum information processing, this property can be exploited in various devices/scenarios, for example to construct input-output registers such as the one described in Fig.15 of Ref. \cite{ronke2011_2}. Because of the `mirroring' property, it is then important to assess the effect of disorder-driven localisation on injection also in sites distant from the chain ends, and in this respect the study of injection at the chain end and centre sites allows us to assess the effect of disorder-driven localisation in the `worst' and `best' case injection scenarios. The excitation number is preserved by the Hamiltonian (\ref{hami}), $[\nonumber{\cal{H}},\sum_{i} |1\rangle \langle 1|_{i}]=0$, even in the presence of disorder, so the system remains in the single excitation subspace of the chain.
The system state at any time can thus be written as

\begin{equation}
|\Psi(t)\rangle = \sum_{i=1}^N c_i(t)|0\rangle_1\dots|1\rangle_i\dots|0\rangle_N
\end{equation}
in the site basis. For a particular realization of diagonal disorder, we have solved the time evolution of the system using both finite step time integration and direct diagonalisation of the full Hamiltonian. For the initial states used in this work, these methods have been found to agree well. For ensemble averages over disorder, we have then solved for a set of size 100 of independent realizations of the disorder. We shall discuss localisation in PST spin chains from both spatial and transport perspectives. We shall analyse also the effect of disorder on the system eigenstates and energy spectrum. We begin with the onset of spatial localisation in PST spin chains, under the effects of diagonal disorder.

\subsection{Spatial localisation}
For infinite one dimensional systems with diagonal disorder, Anderson localisation implies that any initially localised state remains so at all times. For example, in a semi-infinite discrete chain with an excitation started at the end site 1, the localisation condition can be written as

\begin{equation}
\sum_{i=1}^{\infty}|c_i(t)|^2 i < \infty ~\mbox{~at all times t}\; . \label{anderson_loc}
\end{equation}

Similarly, we can express this condition for the excitation started at site $n$ of an infinite chain as

\begin{equation}
\sum_{i=-\infty}^{\infty}|c_i(t)|^2 |i-n| < \infty ~\mbox{~at all times t}\; . \label{anderson_loc2}
\end{equation}

As an example, the semi-infinite case of (\ref{anderson_loc}) is satisfied if the site occupation probabilities follow $|c_i(t)|^2\sim i^{-(2+\delta)}$ for all times and positive $\delta$.
For finite-length chains we will still seek this form of scaling dependence. For an $N$ site chain with occupation probabilities of $p_i = \alpha_N \; i^{-(2+\delta)}$, for relatively large $N$ values, the normalisation factor can be approximated to $\alpha_N = \frac{1} {\zeta (2+\delta)}$, with $\zeta$ being the Riemann Zeta Function (eq. 23.2.1, p. 807 \cite{riemannzeta}). This gives a critical $(\delta=0)$ normalisation of $\alpha_{NCr} = \frac{1} {\zeta(2)}$. For the whole range of $N$ (100 to 1000) in our study,  $\alpha_{NCr}$ is very close to $\alpha_{\infty Cr} \sim 0.6$, so we can plot one critical probability distribution for comparison with all the numerical simulations at different $N$ values, from 100 to 1000.

In Fig.~\ref{endtime2D} we present the effective steady state site occupation probabilities $|c_i|^2$ as a function of $i$ for a range of chain lengths $N=100$ to $1000$, with $E=1$.
To attain an effective steady state, the dynamics is evolved to  $t=5 t_{M}$  when the excitation is injected at site $i=1$ (left panel) and $i=\frac{N}{2}$ (right panel) at $t=0$. Results are then averaged over 100 time steps, from $t=5 t_{M}$ to $t=7 t_{M}$.
A plot of $\alpha_{Cr}\; i^{-2}$ is included in both panels, with $\alpha_{Cr}=\alpha_{\infty Cr}$ in the left, and $\alpha_{Cr}=\alpha_{\infty Cr}/2$ in right panel to account for the double-sided nature of this distribution.
 
For the left panel, comparison with the numerical results shows the range of chains undergoing Anderson localisation according to (\ref{anderson_loc}). Anderson localisation occurs for $N\stackrel{>}{\sim}500$ for injection in the first spin, but not for smaller $N$. Given the approximate straight line behaviour of the data plots, in terms of an analytic approximation to the probabilities given by $p_{i} = \alpha_{N} i^{-(2+\delta)}$, we can comment that $\delta$ exhibits slow variation with $N$. For small $N$, $\delta$ is negative and increasing, crossing the threshold for localisation ($\delta=0$) at about $N \sim 500$.

For the right panel, no real Anderson localisation seems to occur for injection in mid chain and this range of parameters, as the probability distribution does not fall off quickly enough with distance from the injection site.

\begin{figure*} [t]
	\resizebox{1\textwidth}{!}{%
		\includegraphics{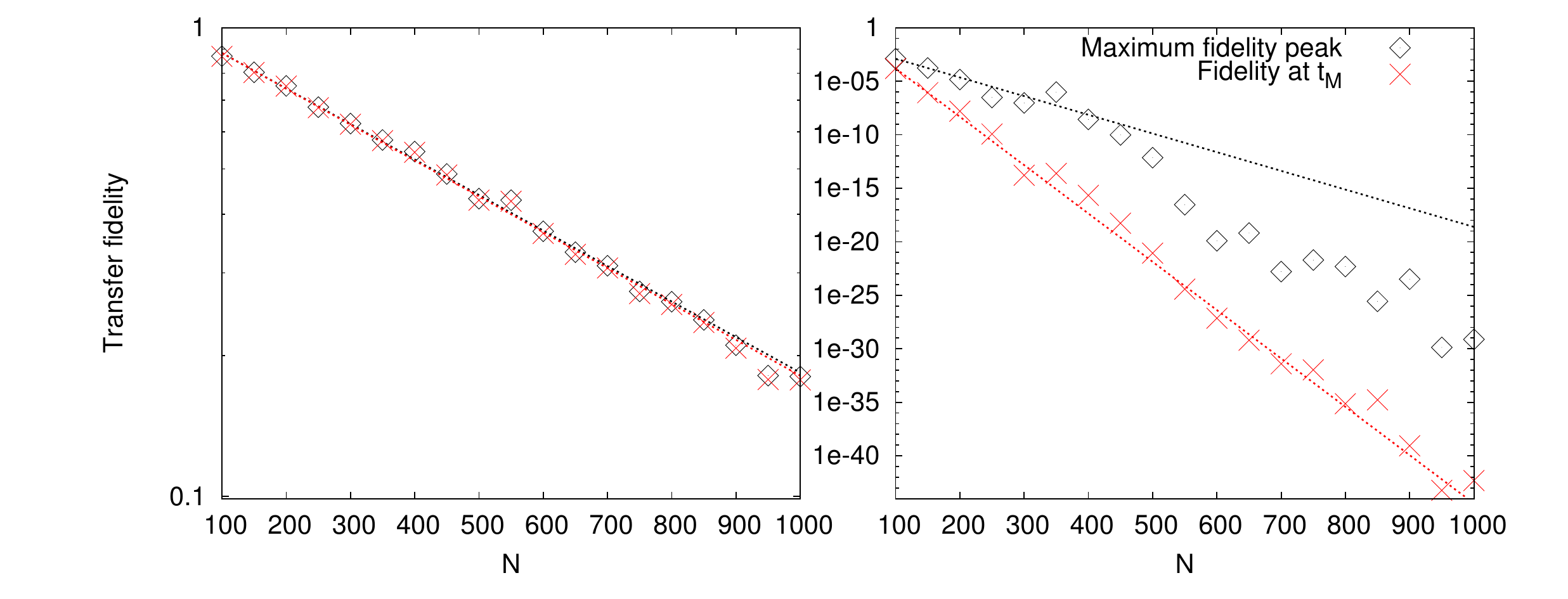}
	}
	\caption{Fidelity of the transferred state in a time window of $4.5t_M$ versus chain length $N$, at exactly $t_M$ (red) and maximum value over $4.5t_M$ (black), $E=0.1$ (left panel) and $E=1.0$ (right panel). We note that both black and red dashed lines are a fit for all values of N.}
	\label{transfer_log}       
\end{figure*}

Nevertheless, for central excitation injection into a PST spin chain the steady state probabilities do still fall off with distance. Furthermore the distribution is essentially independent of $N$ for large $N$, which is to be expected due to the weak $N$-dependence of the spin-spin coupling around the central injection site. In order to more fully capture this `intermediate localisation' behaviour, we consider some further plots. In Fig. \ref{width} we show $n/N$, the fraction of the whole chain over which the steady state probability needs to be summed in order to achieve a total of 0.95, as a function of both disorder strength $E$ and chain length $N$. Here $n$ is defined as the smallest integer such that

\begin{equation}
\frac{1}{N_d}\sum_{d}\frac{1}{N_t}\sum_{t_j}\sum_{i}^{n}|c^{(d)}_i(t_j)|^2 \ge 0.95, \label{n}
\end{equation} 
with $d$ ranging over $N_d=100$ disorder realizations, $t_j$ over the $N_t=100$ discretized time steps from $t=5t_M$ to $t=7t_M$, and $i$ over contiguous sites.

Fig. \ref{width} demonstrates the sharp contrast between the Anderson localisation regime  ($E \sim 1$ and $N \gtrsim 500$) for excitation release at $i=1$, where the chain fraction is close to zero, and the partially localised regime for excitation release at $i=N/2$. 
This is somewhat counterintuitive, as Anderson localisation is normally expected to be independent of initial conditions. However this expectation is correct only when uniform systems are considered, where uniform disorder implies the same local effect on the eigenvalues.
When, as in this case, the unperturbed system is non-uniform, uniform disorder may affect {\it locally} the properties of the system (see also Section \ref{loc}). This is why for disordered PST spin chains it is important to explore injection into locally non equivalent spins. The first and middle spin are the two extremal cases.

Despite this contrast, for excitations released at the chain centre there is some element of localisation, with the fraction decreasing substantially with $N$ for $E \sim 1$. This can be further seen in Fig. \ref{endtime}, where examples of the steady state probability distribution (averaged from $t=5 t_{M}$ to $t=7 t_{M}$) are given for increasing disorder strength $E$. For excitation release at the chain end, the onset of Anderson localisation with increasing $E$ can be clearly seen, with almost all of the probability contained in the first few sites at $E=1$ (bottom panel, left inset). Note that some remnant of PST behaviour is still visible at $E=0.1$, with a small peak in probability at the opposite end of the chain to the injection site. This is removed with increasing $E$, as localisation sets in (bottom panel, right inset). For excitation release at the chain centre, there is clearly an element of localisation, with the probability distribution increasingly peaking at the release location with increasing $E$. For $E=1$ this distribution peak has become essentially independent of $N$, as shown in Fig. \ref{endtime2D}.

\subsection{Localisation and transfer fidelity}\label{sec:3.2}
An alternative and complementary perspective from which to consider Anderson localisation effects is to examine transport.  Our systems of interest without any disorder are by design `perfect wires', that is chains that transport an excitation from one end to the other with perfect fidelity in a time $t_{M}$. Furthermore, modest length $N$ chains with low levels of decoherence (including disorder) exhibit potentially useful robustness against decoherence \cite{ronke2011_1,ronke2011_2}. The transfer maintains high fidelity in this parameter region, which is why PST spin chains are considered to be useful elements for short range quantum communication. Nevertheless, for larger $N$ values there is seen to be exponential damping of the transfer fidelity with $N$, along with Gaussian dependence on the relevant noise amplitude \cite{ronke2011_1,ronke2011_2,chiara2005}. \footnote{We note that in \cite{chiara2005} the diagonal disorder is in units of $J_{0}$ instead of $J_{max}$. However, in units of $J_{max}$ the exponential decay of their Eq.(8) is the same as the damping found in \cite{ronke2011_1,ronke2011_2}, and in particular it displays an exponential damping with $N$. We also note that the scale of the disorder considered in \cite{chiara2005} is such that the Anderson localisation regime is there $not$ accessed for any value of $N$.}

These previous studies have looked at the transfer fidelity at some chosen time, which for example would be $t=t_{M}$ if the objective is perfect quantum communication along a chain. However to link such transport studies to the onset of localisation effects, it is important to examine the fidelity over a range of times to ensure that the maximum transfer fidelity is precisely determined. In fact one contribution to fidelity loss could simply be a shift in the time of an excitation arriving at its destination, rather than a suppression of the arrival happening at all. It is only the latter, and not the former, that is indicative of localisation. To demonstrate suppression of PST consistent with localisation, we have therefore sought the maximum value of the state transfer fidelity over a significant range of time spanning a number of durations of $t_{M}$, and long enough for steady states to be attained when this is a relevant aspect of the behaviour.

In Fig. \ref{avg_max_3D} we show detailed results for the maximum state transfer fidelity attained in a time window of size $4.5 t_{M}$, as a function of both chain length $N$ and disorder strength $E$, for an excitation released at the chain end site $i=1$.  The PST behaviour is clearly visible for all $N$ at zero disorder $E=0$, along with the region of high fidelity for modest $N$ and small $E$ that demonstrates the practical application regime of PST spin chains for short range communication. However, the plot is dominated by a regime of vanishingly small fidelity. This is clearly consistent with previously observed fidelity damping. Given that the plot is of maximum fidelity over a significant time window, this is also clear evidence for the onset of Anderson localisation, complementary to the spatial distribution data given in the previous section.

Further detailed transport data are presented in Fig. \ref{transfer_log}. For weak disorder ($E=0.1$) the maximum transfer fidelity occurs for $t=t_{M}$ and falls off exponentially but weakly with $N$ (left panel), demonstrating practical and usable high fidelities for modest $N \sim 100$. For stronger disorder $E=1.0$ there is rapid (with $N$) exponential fidelity decay (right panel), even for the maximum fidelity over the time window. This demonstrates the strong suppression of PST for a strength of disorder that enables Anderson localisation spatially.

\begin{figure}[t]  
	
	\resizebox{0.5\textwidth}{!}{%
		\includegraphics{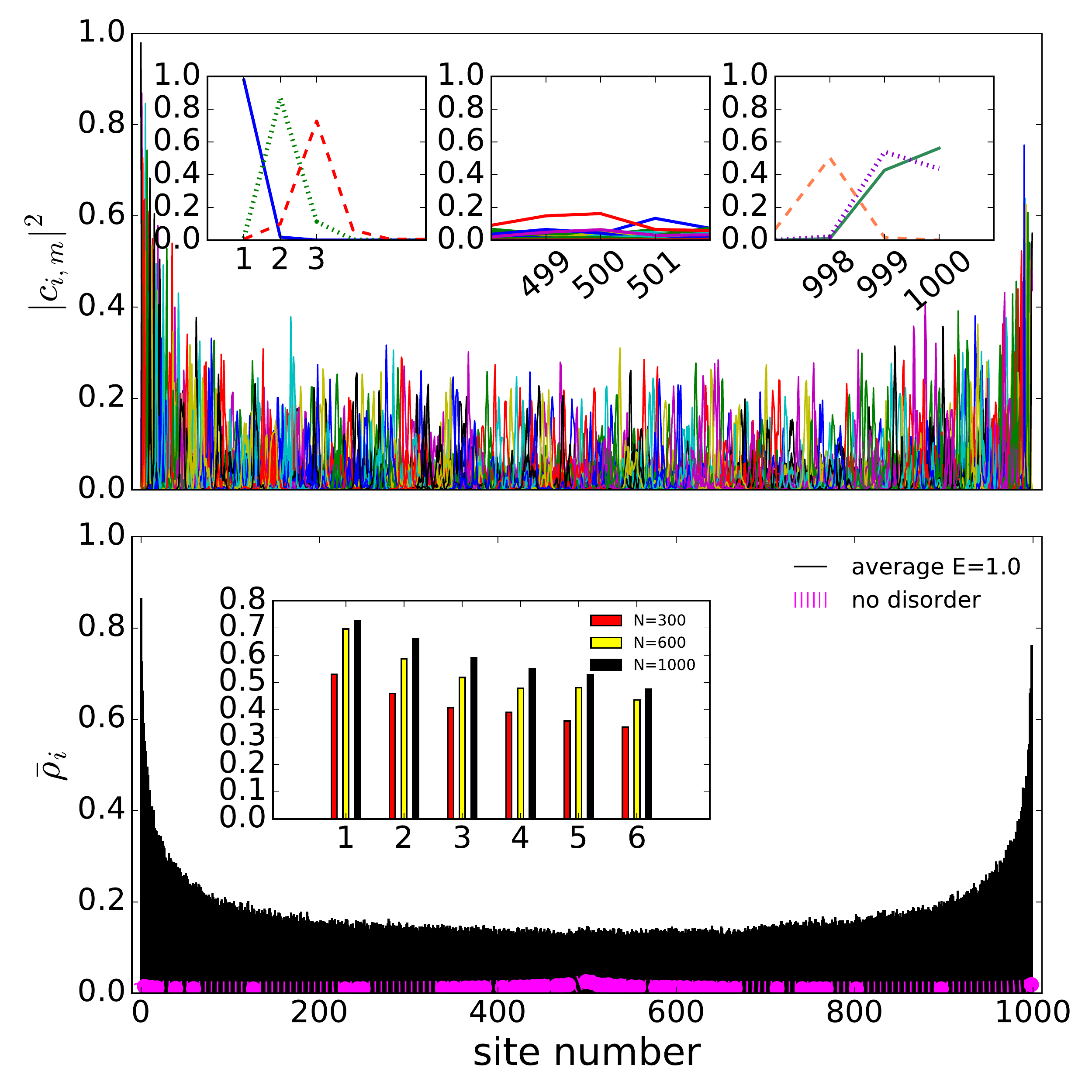}
	}
	\caption{Site occupation probabilities versus site number $i$ for the $N$ eigenstates of the system for one random realization (top panel); maximum occupation probabilities for each site averaged over 100 independent realizations versus site number (bottom panel). Each inset in the top panel shows three eigenstates which peak on the three initial (left inset), middle (center inset) and final (right inset) sites of the chain. The inset in the bottom panel shows the difference between the averaged maximum probabilities at the first six sites for $N$=1000, 600 and 300 (as labeled).}
	\label{eigenvectors}      
\end{figure}

\begin{figure}[t]

	\resizebox{0.5\textwidth}{!}{%
		\includegraphics{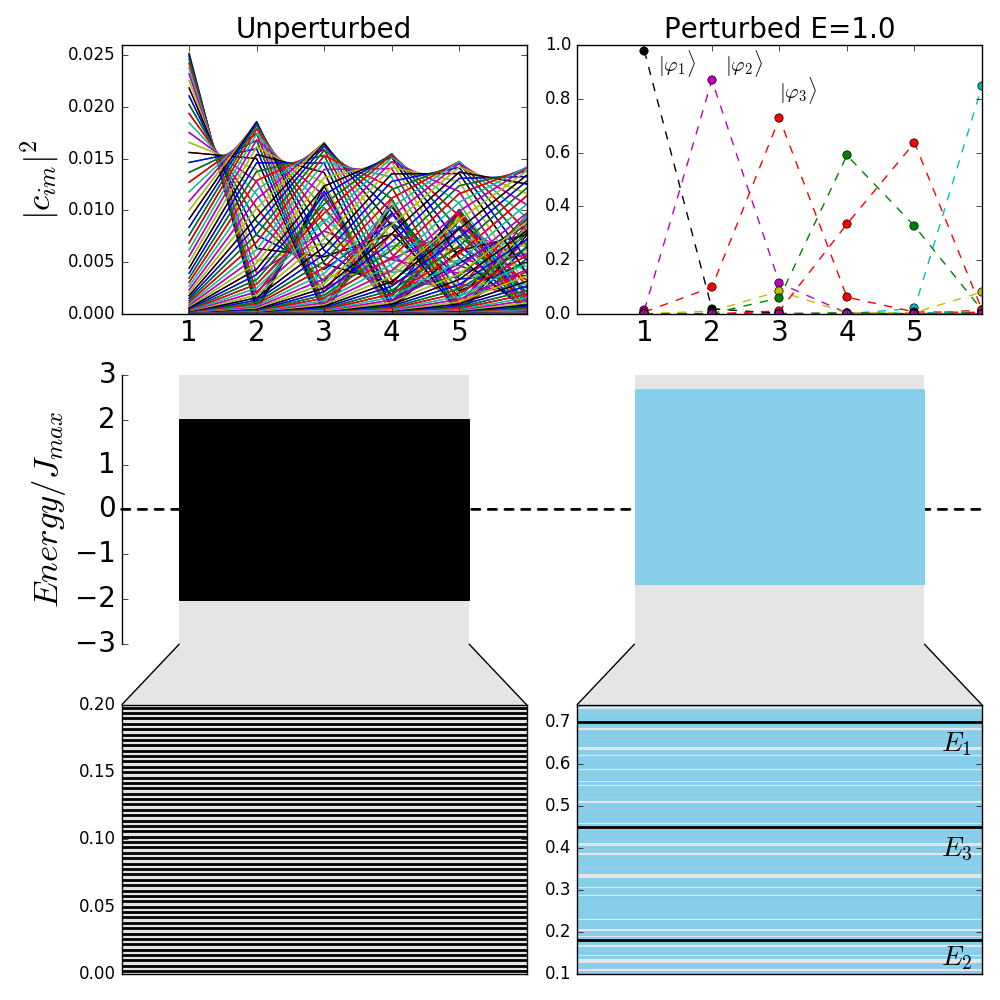}
	}
	\caption{Site occupation probabilities of the eigenstates peaking at the first few sites of an unperturbed (l.h.s) and perturbed $N=1000$ chain with $E=1.0$ (r.h.s) (top panels) and their energy spectra (other panels). Both bottom panels display a zoom to clearly observe the energy splitting, equally spaced for the unperturbed chain and randomly spaced (white gaps) for the perturbed chain. The black lines in the bottom r.h.s. panel indicate the eigenenergies corresponding to the eigenstates $|\varphi_{1}\rangle, |\varphi_{2}\rangle, |\varphi_{3}\rangle$ in the top r.h.s. panel.}
	\label{levels}     
\end{figure}

\subsection{Localisation of the eigenstates}\label{loc}

Signatures of localisation are also apparent from the study of the eigenstates and spectrum of spin chains \cite{zwick2011}. In order to study how the diagonal random disorder affects the eigenstates, $|\varphi_m\rangle$, of the system we obtain the occupation probabilities as function of the site number $i$, $|c_{i,m}|^{2}=|_{i}\langle 1| \varphi_{m} \rangle|^{2}$, with a strength disorder set to $E=1.0$.

The upper panel of  Fig. \ref{eigenvectors} shows these probabilities for a single realization. The left (right) inset demonstrates that the eigenstates which peak at the first (last) three sites of the chain are indeed completely localised within a few sites. This is the expected signature of Anderson localisation for the eigenstates of the Hamiltonian. We further support our findings by considering $\bar\rho_{i}$, the maximum site occupancy probability over all the eigenstates averaged over 100 realizations (average denoted by a bar in the r.h.s. of Eq. \ref{avg}). 

\begin{equation}
\bar{\rho_{i}} \equiv\overline{\max_{m}|\langle i | \varphi_{m} \rangle|^{2}}
\label{avg}
\end{equation} 

This is presented in the lower panel of Fig. \ref{eigenvectors}: here the maximum probability of an eigenstate being in the first (last) sites is very close to unity and increases with the size of the chain (see inset). We contrast this with the corresponding probability distribution for the unperturbed case (pink dashed line profile): in this case all the states are delocalised along the chain and hence the maximum probability of occupying any site in the chain is very small and roughly uniform all along the chain. This delocalisation of all eigenstates is crucial for PST. Injection of an initially localised excitation, for example at the end of the chain, is thus injection of a superposition over many eigenstates, giving raise to the well known PST dynamics. When disorder localises the relevant eigenstates at the chain ends, the same initial injection is then a superposition with far fewer significant amplitudes corresponding to more localised eigenstates (eventually just one, for large disorder).
	
Suppression of transport due to Anderson localisation is explained by combination of eigenstates localisation and splitting of the relevant eigenenergies. The upper panels of Fig.\ref{levels} show the eigenstates localised by disorder on the right contrasted with the delocalised unperturbed states on the left, for the first few sites of the chain. The lower panels show the corresponding energy spectra. For the unperturbed case, the energy levels form a band with an equally spaced distribution of $m$ energy values such that $E_{m}=(N-2m+1)J_{0}$. However, when random disorder is added such that $\epsilon_{i} \ne 0$ and we are in the Anderson localisation regime (considering injection in the first site), the perturbed energy levels are no longer uniformly distributed in the band, and gaps begin to appear (see r.h.s lowest panel of Fig. \ref{levels}). Importantly, the energies corresponding to the eigenstates localised on the few first sites are well separated, as shown by the black lines in the lowest r.h.s panel.

The process of Anderson localisation can be further exemplified by comparing the initial state as injected at site 1 for both the unperturbed and perturbed cases shown in the upper panels of Fig.\ref{levels}. At time $t=0$, the initial state will be a superposition of all the non-vanishing eigenstates at site 1 and its energy will be the corresponding linear combination of eigenenergies. For the unperturbed case, these eigenstates will be many, and by inspection and by considering that the energy band is very dense, we may deduce that the energy from the corresponding linear combination of eigenenergies will not be very different when moving from the first to the next sites (this is in fact corroborated by PST). 

However, for the disordered chain, the state as injected in site 1 is approximately,

\begin{equation}
	|\Psi_{inj}\rangle=a_{11}|\varphi_{1}\rangle+a_{12}|\varphi_{2}\rangle+a_{13}|\varphi_{3}\rangle ,
	\label{aprox}
\end{equation}
and will be dominated by approximately one eigenstate, $|\varphi_{1}\rangle$ (see Fig.\ref{levels}, top r.h.s. panel). 
Its energy will then be
\begin{equation}
	\langle \Psi_{inj}| \nonumber{\cal{H}}|\Psi_{inj}\rangle=\sum_{i} \langle \Psi_{inj}|a_{1i}E_{i}|\varphi_{i}\rangle ,
	\label{en}
\end{equation}
and we can approximate $\langle\Psi_{inj}|\nonumber{\cal{H}}|\Psi_{inj}\rangle \approx E_{1}$ as $|a_{11}|^2 \gg |a_{12}|^2,|a_{13}|^2$ (see upper r.h.s panel of Fig.\ref{levels}). 

Bearing in mind that the dynamical evolution of our Hamiltonian will conserve the energy, and having shown that our initial state sits on the $E_{1}$ energy level, we can conclude in addition to the eigenstate localisation argument that, because the eigenstates peaking at the nearby sites have energies far apart from the initial one (see Fig.\ref{levels}, bottom-right panel), the state transfer will be strongly diminished. Therefore the presence of Anderson localisation for this specific case is indeed supported by both eigenstate and eigenenergy arguments.

\subsection{Considerations on heat transport}\label{sec:3.4}

We have restricted our localisation studies in this paper to the single excitation subspace: even in the presence of disorder, the system Hamiltonian conserves excitation number, and this subspace enables modelling relevant for the Anderson localisation scenario. Our results allow us also to comment on heat transport across the chain, at least for the case in which a heat reservoir is put in touch with one of the end spins in such a way that this spin is placed in a thermal mixture (of zero and one excitation). If the chain is a PST-type chain, we know there is the potential for perfect transmission of this state across the chain. However, our current study of the suppression of transfer fidelity as a function of growing disorder (section \ref{sec:3.2}) shows how this heat transport across the chain is progressively suppressed. To study thermal states of multiple spins and/or also broader energy transport scenarios, inclusion of multiple excitation subspaces would be necessary, which goes beyond the scope of the present work.

\section{Conclusions}

In this paper we have investigated the phenomenon of localisation in one-dimensional PST spin chains. For low disorder these chains are known to exhibit some robustness in their transport properties, which is why they are of interest for quantum information transport. However, our investigations have shown how transport is suppressed for medium levels of disorder ($E \sim 1$) and as a function of chain length $N$, due to the spatial localisation of a locally injected excitation. Detailed investigations of the steady state spatial probability distributions for injected excitations reveal different localisation effects for injection at a chain end and at the centre. We explain such differences with the PST coupling scheme of Eq.(\ref{PST}) which gives different site dependence of the spin-spin coupling for these different chain regions. Excitations at the chain centre demonstrate an element of localisation for disorder at strength $E \sim 1$, whereas excitations at the chain end exhibit genuine Anderson localisation for this level of disorder and chains longer than $N \sim 500$. These different regimes of localisation are also reflected in the localisation of the system eigenstates, with eigenstates localised over few sites at the beginning and end of the chain (Anderson localisation) for $N \gtrsim 500$ and $E \sim 1$. These localised states have energies that sit far apart in the energy spectrum, further supporting the lack of hopping between stated sites. Our work thus provides another interesting physical system that exhibits localisation phenomena. Future work will examine the potential of controlled disorder in these systems being used as a tool for manipulating spin chain properties and behaviour. Also, although computationally more demanding, there is further motivation for future studies to include subspaces with higher number of excitations. There is growing interest in the concept of many-body localisation, and the potential link between correlations and the onset of such many-body localisation \cite{goold2015}. It would therefore be interesting to examine spin correlations in multiple-excitation subspaces as a function of increasing disorder, exploring this potential link in spin chains.

\section*{Acknowledgements}

RR was supported by EPSRC-GB and Hewlett-Packard.\\
MPE was supported by the University of York.

\bibliographystyle{unsrt}
\bibliography{papers_ire}

\end{document}